\documentclass[final,3p,times,twocolumn]{elsarticle}
\usepackage{color}
\usepackage[numbers]{natbib}

\usepackage{booktabs} 
\usepackage{tabularx}
\usepackage{caption} 
\usepackage{siunitx}
\usepackage{amsmath}
\usepackage[version=4]{mhchem}
\definecolor{ColorName1}{rgb}{0.0, 0.0, 0.0}

\begin{document}
\title{Optical coherence and hyperfine structure of the ${}^7\mathrm{F}_0\leftrightarrow{}^5\mathrm{D}_0$ transition in $\rm Eu^{3+}$:$\mathrm{CaWO_4}$ }   

\author{Xiantong An$^{a,c}$}
\author{Weiye Sun$^{a,c}$}

\author{Zhehao Xu$^a$}
\author{Wanting Xiao$^a$}
\author{Miaomiao Ren$^a$}
\author{Mucheng Guo$^a$}

\author{Shuping Liu$^{b,\ast}$}
\ead{liushuping@iqasz.cn}

\author{Fudong Wang$^{a,b,\ast}$}
\ead{fdwang.phys@foxmail.com}

\author{Manjin Zhong$^{a,b,\ast}$}
\ead{manjin.zhong@gmail.com}

\affiliation{Shenzhen Institute for Quantum Science and Engineering, Southern University of Science and Technology, Shenzhen 518055, China}

\affiliation{International Quantum Academy, and Shenzhen Branch, Hefei National Laboratory, Shenzhen, 518048, China}

\affiliation{These authors contributed equally to this work}

 \cortext[1]{Corresponding authors}

\begin{abstract}
Rare-earth ions doped in crystals with low nuclear-spin densities are highly promising candidates for quantum technology applications. In this study, we investigated the spectroscopic properties of the $^7\mathrm{F}_0\leftrightarrow^5\mathrm{D}_0$ optical and the hyperfine transitions of $\rm Eu^{3+}$ ions in a CaWO$_4$ crystal, where the nuclear spin arises solely from the $\rm^{183}W$ isotope, with a natural abundance of 14 \%. At a temperature of 3 K, we experimentally identified four distinct crystal field environments for $\rm Eu^{3+}$ ions in a 0.1 at.\% $\rm Eu^{3+}$ doped CaWO$_4$ crystal. The optical coherence properties of Eu$^{3+}$ ions in these environments were characterized. Additionally, we resolved the hyperfine structures in the $^7\mathrm{F}_0$ ground state and $^5\mathrm{D}_0$ excited state, and determined the $^7\mathrm{F}_0$ ground state lifetimes using spectral hole burning techniques. These findings highlight the significant potential of $\mathrm{Eu^{3+}}$:$\mathrm{CaWO_4}$ for optical quantum memory applications.
\end{abstract}

\begin{keyword}
$\mathrm{Eu^{3+}}$:$\mathrm{CaWO_4}$ \sep high resolution spectroscopy \sep optical decoherence \sep hyperfine structure \sep quantum memory
\end{keyword}

\maketitle

\section{Introduction}

Rare-earth-ion-doped crystals, renowned for their exceptional optical and spin coherence, have emerged as key materials in quantum information science. These systems enable long-lived quantum state storage and manipulation, with applications  \textcolor{ColorName1}{spanning from quantum communication to frequency conversion} between optical and microwave regimes \cite{probst2015microwave, williamson2014magneto, fernandez2015coherent, Rai12}. To achieve optimal performance, host crystals must exhibit minimal magnetic noise to preserve ion coherence. While significant progress has been made using existing host materials\cite{ PhysRevLett.95.030506, Rancic2018, Ma2021, PRXQuantum.6.010302, zhong2015optically, zhong2019quantum}, further advancements require exploring alternative hosts with inherently low nuclear spin noise or "nuclear spin-free" properties \cite{PRXQuantum.4.010323, d9217a51}.

Calcium tungstate ($\mathrm{CaWO_4}$) offers unique advantages due to its low nuclear spin noise. Its natural isotopic composition minimizes magnetic noise, as only $\mathrm{{}^{183}W}$ contributes appreciably, with a natural abundance of 14\% and a low magnetic dipole moment of \num{0.117}$\mu_N$ \cite{bruno2014crc, GARBACZ20177}. These characteristics position $\mathrm{CaWO_4}$ as an ideal platform for studying rare-earth ion coherence and developing quantum devices \cite{le2021twenty, uysal2024spin}.

Notable progress has been made in doping $\mathrm{CaWO_4}$ with ions like $\mathrm{Er^{3+}}$, \textcolor{ColorName1}{$\mathrm{Gd^{3+}}$}, $\mathrm{Yb^{3+}}$, $\mathrm{Ce^{3+}}$, and $\mathrm{Pr^{3+}}$ \textcolor{ColorName1}{\cite{rakhmatullin2009coherent, yang2023spectroscopic, 10.1063/5.0224102}}, enabling research on spin coherence and quantum storage. However, studies on $\mathrm{Eu^{3+}}$ doping remain limited. As a rare earth ion with intricate hyperfine structure and favorable quantum properties, $\mathrm{Eu^{3+}}$ is particularly suited for optical quantum memory development and precise quantum state control. Despite its potential, prior research has primarily explored high-temperature spectral properties of $\mathrm{Eu^{3+}}$:$\mathrm{CaWO_4}$ \cite{BAI2020117351, ZHANG2018115}, leaving its coherence and low-temperature behavior largely unexplored.

This work addresses these gaps by investigating $\mathrm{Eu^{3+}}$:$\mathrm{CaWO_4}$ at liquid helium temperatures. Detailed studies \textcolor{ColorName1}{including} the $^7\mathrm{F}_0\leftrightarrow^5\mathrm{D}_0$ optical transition, coherence times, and hyperfine structure are presented. Additionally, the lifetimes of the $^5\mathrm{D}_0$ excited state and $^7\mathrm{F}_0$ ground hyperfine state are measured. The results highlight the \textcolor{ColorName1}{great potential of the $\mathrm{Eu^{3+}}$:$\mathrm{CaWO_4}$ crystal for quantum memory and spin-wave-based} applications in quantum technologies.

\section{Experiments}
The $\mathrm{Eu^{3+}}$:$\mathrm{CaWO_4}$ single crystal used in this study was grown by Czochralski method and then post-treated at 1000 $^\circ$C for 8 h \textcolor{ColorName1}{under} air atmosphere. It has a tetragonal structure with space group $I4_1/a$ and unit cell parameters of a = b = \SI{5.24}{\angstrom} and c=\SI{11.38}{\angstrom} \cite{mims1967local}. Due to the similar ionic radii of $\mathrm{Eu^{3+}}$ (\SI{1.06}{\angstrom}) and $\mathrm{Ca^{2+}}$ (\SI{1.12}{\angstrom}) \cite{Felinto2017Photoluminescence}, $\mathrm{Eu^{3+}}$ substitute $\mathrm{Ca^{2+}}$ sites \textcolor{ColorName1}{readily within an 8-fold oxygen coordination environment} (distorted dodecahedron, point group S$_4$). In this work, the $\mathrm{Eu^{3+}}$ was doped with a concentration of 0.1 at.\%. After being cut and polished along the $a\times b\times c$-axis with \numproduct{5 x 6 x 7} \unit{\mm^3} dimension, \textcolor{ColorName1}{the crystal was mounted into a $\mathrm{CryoAdvance^{TM}}$ 100 cryostat (Montana Instruments) and cooled to 3 K for further spectroscopic measurement.} 

The inhomogeneous linewidth ($\Gamma\mathrm{_{inh}}$) of the $^7$F$_0\rightarrow^5$D$_0$ transition (\textcolor{ColorName1}{$\lambda_{vac} = 580.8$ nm}) was recorded by monitoring the $^5$D$_0\rightarrow^7$F$_2$ fluorescence \textcolor{ColorName1}{($\lambda_{vac} = 611$ nm) during frequency scans of a single-pass excitation laser beam propagating along the crystallographic $c$-axis. The laser system consists of a Toptica seed source, a Precilasers fiber amplifier, and a frequency-doubling stage. The laser frequency was measured with a HighFinesse WS6-200 wavemeter, which was calibrated via saturated absorption spectroscopy of $^{85}\mathrm{Rb}$ atoms.} The excitation power was \SI{2}{\milli\watt} and the fluorescence signals were collected with a long-pass filter (610 nm cut-off wavelength) and a photomultiplier tube (HAMAMATSU, H10721-20). The excited state lifetime was measured by recording the decay of fluorescence. The optical coherence property and hyperfine structure of $^7$F$_0\rightarrow^5$D$_0$ transition was measured by two pulse photon echo and spectral hole burning \textcolor{ColorName1}{(SHB)} methods, respectively. To enhance the absorption of $\mathrm{Eu^{3+}}$ in $\mathrm{CaWO_4}$ during the measurements, a pair of planar mirrors were employed to allow for a four-pass of the laser through the crystal.

\section{Results and discussion} 

\subsection{Optical excitation characterization}

\begin{figure*}[htbp]
    \centering
    \begin{minipage}[t]{0.14\textwidth}
        \centering
	\vspace{1pt}
        \includegraphics[width=\linewidth]{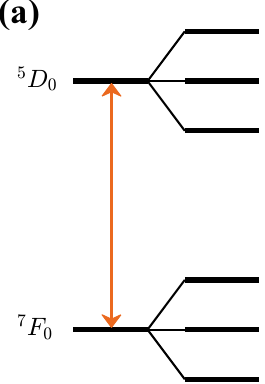}
    \end{minipage}
    \hfill
    \begin{minipage}[t]{0.8\textwidth}
        \centering
		\vspace{0pt}
        \includegraphics[width=\linewidth]{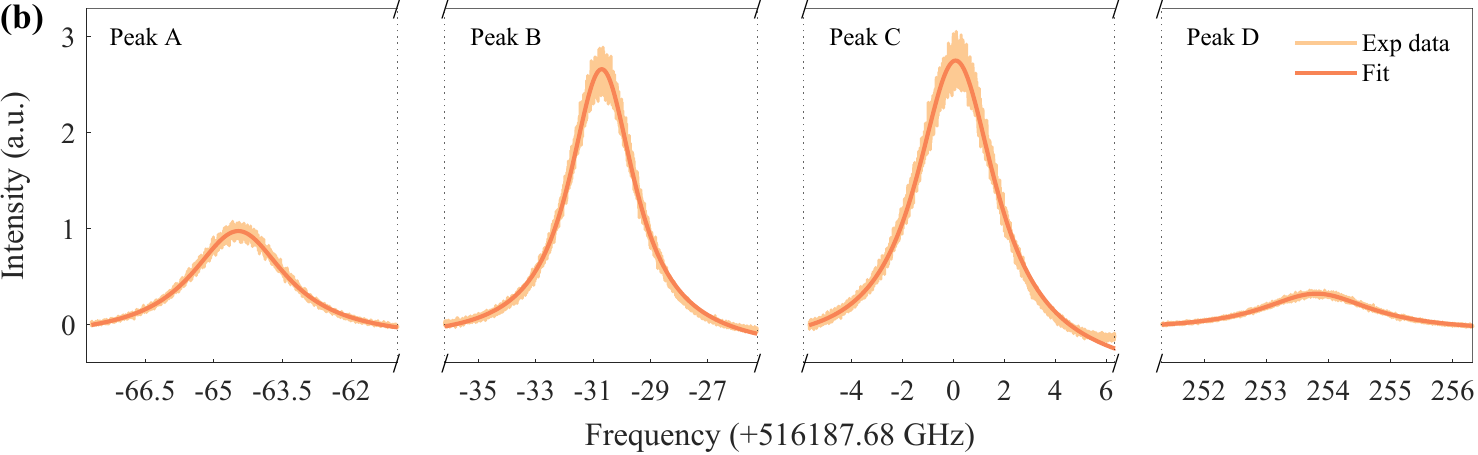}
    \end{minipage}
    \caption{(a) Hyperfine structures of the ${}^7\mathrm{F}_0\leftrightarrow{}^5\mathrm{D}_0$ transition. (b) Fluorescence spectra in which $\mathrm{k} \parallel \mathrm{c}$, fitted through Lorentzian \textcolor{ColorName1}{function}, with the results shown in Table \ref{Table:1}.}
    \label{Fig:1}
\end{figure*}

\begin{table*}[h]
	\fontsize{12pt}{12pt}\selectfont
	\caption{\textcolor{ColorName1}{The} spectroscopic properties of the ${}^7\mathrm{F}_0 \leftrightarrow {}^5\mathrm{D}_0$ optical and hyperfine transitions for different peaks. \textcolor{ColorName1}{The center frequency of Peak C is \textcolor{ColorName1}{$516187.75 \pm 0.02$ GHz ($580.78 \pm 0.02$ nm)}, while the $f_c$ values provided for the other three peaks are shifted relative to the center frequency of the Peak C.}}
	\label{table1}
	\centering
	\resizebox{16.5cm}{!}
	{\begin{tabular}{ccccccccc} 
			\toprule
			Peaks & \multicolumn{1}{c}{$f_{\text{c}}$ (GHz)} & \multicolumn{1}{c}{FWHM (GHz)} & \multicolumn{1}{c}{OD}   & $\mathrm{T_2 \, (\mu s)}$ & $\alpha_{TPR}$ $(*10^{-6} \, \text{Hz} / \text{K}^7)$  & $\textcolor{ColorName1}{\Gamma_{hom0}} \, (\text{kHz})$ & $\mathrm{T_1}\, (\mu s)$  & $\mathrm{T_{1,spin}\, (s)}$\\
			\midrule
			Peak A & \textcolor{ColorName1}{-64.54} & \textcolor{ColorName1}{$2.54 \pm 0.02$} & 0.033 & $ \textcolor{ColorName1}{114 \pm 7}  $ & $ \textcolor{ColorName1}{4.07 \pm 0.24} $ & $ \textcolor{ColorName1}{2.74 \pm 0.08} $ & $ \textcolor{ColorName1}{562 \pm 7} $ & $ \textcolor{ColorName1}{688 \pm 1} $\\
			Peak B & \textcolor{ColorName1}{-30.78} & \textcolor{ColorName1}{$2.80 \pm 0.04$} & 0.078 & $ \textcolor{ColorName1}{114 \pm 7}  $ & $ \textcolor{ColorName1}{4.35 \pm 0.08} $ & $ \textcolor{ColorName1}{2.82 \pm 0.03} $ & $ \textcolor{ColorName1}{597 \pm 2} $ & $ \textcolor{ColorName1}{1161 \pm 3}$ \\
			Peak C & 0.00 & \textcolor{ColorName1}{$3.85 \pm 0.07$}  & 0.083 & $ \textcolor{ColorName1}{105 \pm 8} $ & $ \textcolor{ColorName1}{10.26 \pm 0.21} $ & $ \textcolor{ColorName1}{2.82 \pm 0.07} $ & $ \textcolor{ColorName1}{581 \pm 3} $ & $ \textcolor{ColorName1}{779 \pm 9} $\\
			Peak D & \textcolor{ColorName1}{253.72} &  \textcolor{ColorName1}{$2.24 \pm 0.01$} & 0.015 & $ \textcolor{ColorName1}{111 \pm 11}  $ & $ \textcolor{ColorName1}{3.61\pm 0.21} $ & $ \textcolor{ColorName1}{3.12 \pm 0.07} $ & $ \textcolor{ColorName1}{518 \pm 3} $ & $ \textcolor{ColorName1}{1215 \pm 4} $\\
			\bottomrule
	\end{tabular}}
	\label{Table:1}
\end{table*}

Fig.\ref{Fig:1} shows the energy levels and inhomogeneous line of the ${}^7\mathrm{F}_0\leftrightarrow{}^5\mathrm{D}_0$ transition. \textcolor{ColorName1}{Unexpectedly,  four excitation peaks are observed near 580.8} nm, with the center frequency spanning \textcolor{ColorName1}{a range of} over 300 GHz. All the lineshapes \textcolor{ColorName1}{of all four peaks} could be described by \textcolor{ColorName1}{Lorentzian} functions, giving a inhomogeneous linewidth ($\Gamma\mathrm{_{inh}}$, full width at half maximum) of \textcolor{ColorName1}{$2.54 \pm 0.02$} GHz, \textcolor{ColorName1}{$2.80 \pm 0.04$} GHz, \textcolor{ColorName1}{$3.85 \pm 0.07$} GHz and \textcolor{ColorName1}{$2.24 \pm 0.01$ GHz for the labeled Peak A, B, C and D, respectively.} This indicates that there are multiple \textcolor{ColorName1}{crystallographic} environments \textcolor{ColorName1}{for $\mathrm{Eu^{3+}}$ ions in the crystal} \cite{becker2024spectroscopicinvestigationsmultipleenvironments}. Absorption spectra of the ${}^7\mathrm{F}_0\leftrightarrow{}^5\mathrm{D}_0$ transition \textcolor{ColorName1}{were also measured using a four-pass configuration, yielding consistent results.} \textcolor{ColorName1}{Notably, Peak B, located close to Peak C, exhibits an intensity similar to Peak C but significantly stronger than Peak A and D}. The center frequency, $\Gamma\mathrm{_{inh}}$ and the optical depth (OD) from absorption measurement for the four peaks are listed in Table \ref{Table:1}. Due to the low doping concentration, an absorption coefficient of \SI{0.12}{\per\centi\meter} and a maximum OD of 0.083 is resolved for Peak C.

\begin{figure}
	\centering
	\includegraphics[width=.5\textwidth]{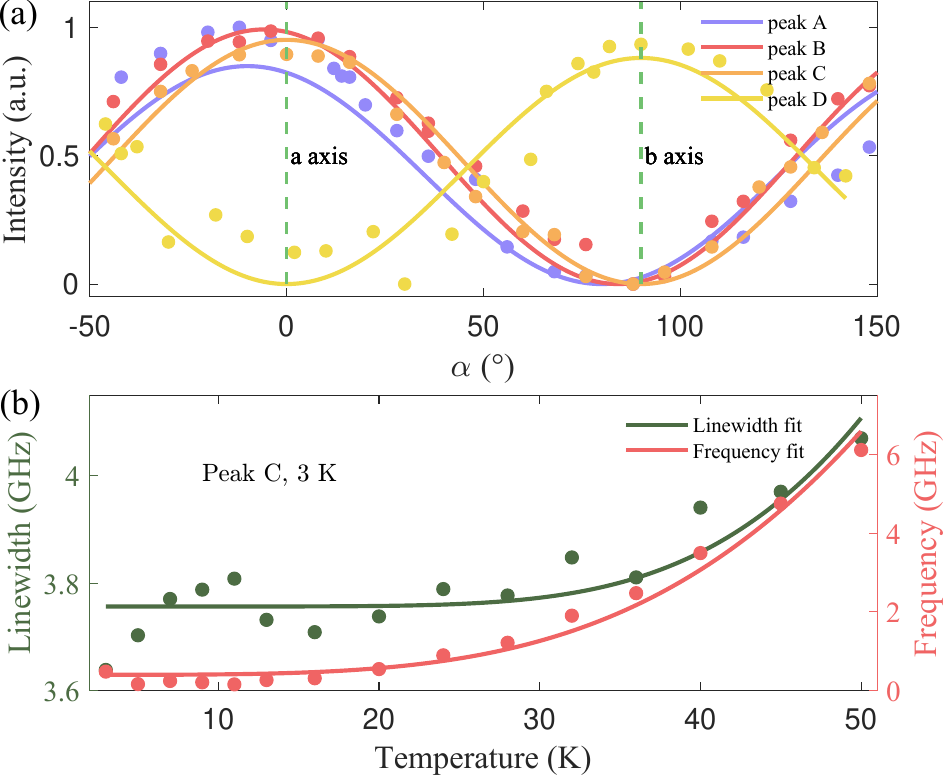}
	\caption{(a) Polarization characteristics of the excitation spectra, in which $\mathrm{k} \parallel c$, have a period of $\pi$. (b) The trend of linewidth (left scale) and the transition of ${}^7\mathrm{F}_0\leftrightarrow{}^5\mathrm{D}_0$ (right scale) in the Peak C as a function of temperature, where the right scale represent the frequency offset relative to \textcolor{ColorName1}{516187.75} GHz. And scatter points represent the original data and the colored curves represent the fitting results.}
	\label{Fig:2}
\end{figure}

We then studied the polarization dependence of the excitation intensity for the four peaks. \textcolor{ColorName1}{As shown in }Fig.\ref{Fig:2}a, \textcolor{ColorName1}{the results indicate that} \textcolor{ColorName1}{Peak} A, B, and C \textcolor{ColorName1}{shared a similar pattern}, \textcolor{ColorName1}{whereas Peak D} \textcolor{ColorName1}{displays} an inverse trend with a period of $\pi$. The dependence of the intensity on the laser polarization angle ($\alpha$) with respect to the $a$-axis is consistent with the expression \cite{afzelius2010efficient}:
\begin{equation}
	I(\alpha)=[\cos^2(\alpha)e^{-d_\parallel}+\sin^2(\alpha)e^{-d_\perp}]I_0:=e^{-d(\alpha)}I_0
\end{equation}
where $I_0$ is input intensity, ${d_\parallel}$ and ${d_\perp}$ are the effective depths for the polarization parallel and perpendicular to the $a$-axis, respectively.

The temperature dependence of the linewidth and center frequency of the ${}^7\mathrm{F}_0\leftrightarrow{}^5\mathrm{D}_0$ transition for \textcolor{ColorName1}{Peak} C is shown in Fig.\ref{Fig:2}b. \textcolor{ColorName1}{The figure clearly demonstrates that} inhomogeneous line broadening and frequency shift of the transition \textcolor{ColorName1}{occur as the} temperature \textcolor{ColorName1}{increases} from 3 K to 50 K. Taking into account of the two-phonon Raman processes \cite{mccumber1963linewidth}, \textcolor{ColorName1}{we fitted the experimental data using the following functions}:
\begin{equation}
	 \Gamma(T)=\Gamma_0+\alpha\left(\frac{T}{\Theta_D}\right)^7 \int_0^{\frac{\theta_D}{T}} \frac{x^6 e^x}{\left(e^x-1\right)^2} \mathrm{~d} x 
\end{equation}
\begin{equation}
	 v(T)=v_0+\bar{\alpha}\left(\frac{T}{\Theta_D}\right)^4 \int_0^{\frac{\Theta_D}{T}} \frac{x^3}{e^x-1} \mathrm{~d} x 
\end{equation}
where T is the temperature, \textcolor{ColorName1}{$\Theta_D$ =355 K is the Debye temperature of $\mathrm{CaWO_4}$ \cite{PhysRevB.70.214306}}, and $\alpha$, $\bar{\alpha}$ are parameters associated with phonon scattering. As shown in Fig.\ref{Fig:2}b, the expression fits well with experimental data and gives: \textcolor{ColorName1}{$\nu_0$ = 516188.10 $\pm$ 0.18 GHz, $\bar{\alpha}$ = (2.60 $\pm$ 0.22)$\times10^3$ GHz, $\Gamma_0$ = 3.76 $\pm$ 0.03 GHz, $\alpha$= (0.76 $\pm$ 0.24)$\times10^3$ GHz}. At temperatures below 10 K, phononic influences on optical transitions and inhomogeneous linewidth are very limited and can be neglected.

\subsection{Optical coherence properties}

\begin{figure}[h]
	\centering
	\includegraphics[width=.48\textwidth]{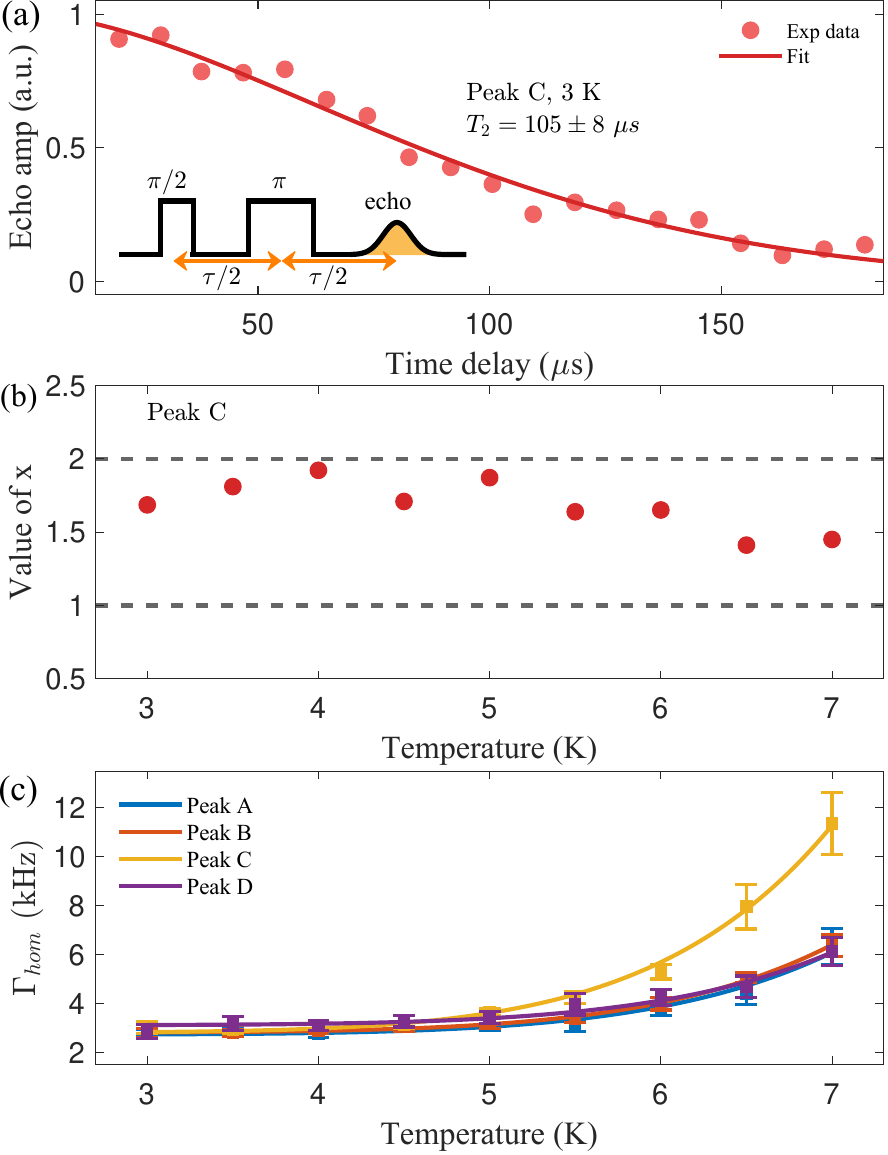}
	\caption{\textcolor{ColorName1}{(a) Photon echo decay of the Peak C and experimental pulse sequence. Considering that the coherence times of four peaks are similar, this paper presents only the photon echo decay of the Peak C and provides the fitting results for the coherence times of four peaks, as shown in Table \ref{Table:1}. (b) Values of x with temperature for the Peak C fitted by Equation 4. (c) The variation of homogenous linewidth with temperature, where the error bars represent the original experimental data and the colored curves are the fitting curves.} }
	\label{Fig:3}
\end{figure}

We investigated the optical coherence \textcolor{ColorName1}{time} (T$_2$) of the ${}^7\mathrm{F}_0\leftrightarrow{}^5\mathrm{D}_0$ transition by performing two-pulse photon echo experiments at a temperature of 3 K \cite{kurnit1964observation}. \textcolor{ColorName1}{The laser propagated along the $a$-axis with a power of 30 mW during the measurement.} By optimizing the exciting $\pi/2$ and rephasing $\pi$ pulses in the sequence to be 3 $\mu$s and 6 $\mu$s, respectively, the photon echo \textcolor{ColorName1}{signals were} detected by using heterodyne detection accompanied with a fast Fourier transform (FFT). The laser pulse sequences and the typical echo decay curves measured \textcolor{ColorName1}{in Peak} C are shown in Fig.\ref{Fig:3}a. Invoking the echo relaxation model proposed by \textcolor{ColorName1}{Mims \cite{PhysRev.168.370}}, the optical T$_2$ was derived from the fitting of the echo amplitude $A$ against the total evolution time $\tau$ in the form of
\begin{equation}
	A=A_0\cdot {\rm exp}[-(\tau/T_2)^x] \label{eq1}
\end{equation}
where x is the parameter describing spectral diffusion. \textcolor{ColorName1}{As shown in Fig.3b, the fitting gives x $\in[1, 2]$} for all four peaks, suggesting prominent spectral diffusion occurs during the time scale of measurements. \textcolor{ColorName1}{The fitting resulted in optical T$_2$ of $114\pm 7~\mu$s, 114$\pm$7 $\mu$s, $105\pm 8 ~\mu$s and $111\pm 11~\mu$s} at 3 K for Peak A, B, C and D, respectively. 

The temperature dependence of homogeneous linewidth ($\Gamma_\mathrm{h}$, $\Gamma_\mathrm{h}$=1/($\pi T_2$)) for the four peaks is illuminated in Fig.\ref{Fig:3}c. It is observed that the $\Gamma_\mathrm{h}$ values of all peaks start to increase rapidly \textcolor{ColorName1}{above 6 K, which is attributed to the }two-phonon Raman (TPR) scattering process \cite{mccumber1963linewidth}. To quantify the decoherence dynamics of the four peaks, we fitted the experimental data by the expression \cite{konz2003temperature}:
\begin{equation}
	\Gamma_{hom}=\textcolor{ColorName1}{\Gamma_{hom0}}+\alpha_{\mathrm{TPR}}T^7  
\end{equation}
where $\textcolor{ColorName1}{\Gamma_{hom0}}$ is the residual homogeneous broadening, $\alpha_{\mathrm{TPR}}$ is the two-phonon Raman coupling coefficient. The fitting results are listed in Table \ref{Table:1}, where \textcolor{ColorName1}{Peak} C displays the largest $\alpha_{\mathrm{TPR}}$ value compared to the other three peaks, \textcolor{ColorName1}{consistent with its shortest T$_2$}.

\subsection{Hyperfine structures}

\begin{figure*}
	\centering
	\includegraphics[width=1.0\textwidth]{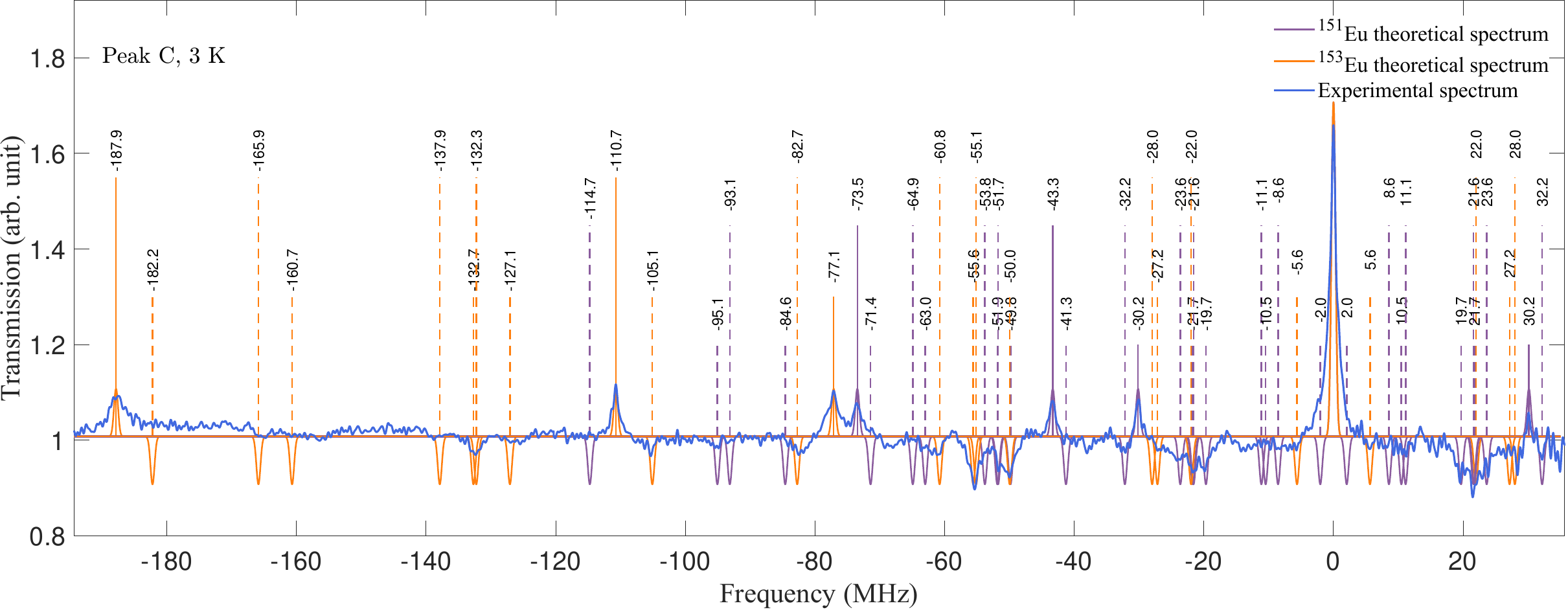}
	\caption{In zero magnetic field, the schematic of the single-frequency \textcolor{ColorName1}{SHB} for Peak C is shown, where the blue line represents the experimentally obtained spectrum, and the purple and orange lines indicate the positions of the theoretical single-frequency \textcolor{ColorName1}{SHB}  for $^{151}\mathrm{Eu}$ and $^{153}\mathrm{Eu}$, respectively. Here, only the frequency positions are valid, and the intensities are not valid. }
	\label{Fig:4}
\end{figure*}

\textcolor{ColorName1}{Europium has two isotopes with approximately equal abundance, $^{151}\mathrm{Eu}$ and $^{153}\mathrm{Eu}$, both have nuclear spin of 5/2.} In this study, we first investigated the hyperfine energy levels of \textcolor{ColorName1}{these} two isotopes located at four peaks using single-frequency \textcolor{ColorName1}{SHB} \cite{babbitt1989optical,silversmith1992spectral,nilsson2004hole,hastings2008spectral,lauritzen2012spectroscopic}.
To achieve \textcolor{ColorName1}{high-precision spectral measurements}, we employed the Pound-Drever-Hall (PDH) locking technique to stabilize the laser \textcolor{ColorName1}{to} an ultrastable optical cavity \cite{taubman2013optical}.
\textcolor{ColorName1}{Frequency scanning was implemented via an acousto-optic modulator (AOM) driven by an arbitrary waveform generator (AWG). Given the high resolution of the AWG, the overall measurement accuracy is fundamentally limited by  the laser frequency locking precision, which is \SI{100}{kHz}.}
 To enhance the absorption, we used \textcolor{ColorName1}{a pair} of mirrors to enable the laser to pass through the sample four times during the \textcolor{ColorName1}{SHB} experiments for \textcolor{ColorName1}{Peak} A, B and C and five times for \textcolor{ColorName1}{Peak} D. The \textcolor{ColorName1}{laser beam propagated} along $c$-axis of the crystal \textcolor{ColorName1}{, with powers of 12 mW for the burn pulse and 5 mW for the probe pulse.}

As described in Fig.\ref{Fig:1}a, the hyperfine energy level structure of $\mathrm{Eu^{3+}}$ ions \textcolor{ColorName1}{allows} for the presence of 1 central hole, 6 side-holes, and 42 anti-holes during the single-frequency \textcolor{ColorName1}{SHB} experiments \cite{sharma2023photon}. By analyzing the hole spectra presented in Fig.\ref{Fig:4} and leveraging our existing knowledge, we identified the hyperfine structures of the europium ions in \textcolor{ColorName1}{Peak} C. Specifically, the $^{151}\mathrm{Eu}^{3+}$ ions exhibit excited state hyperfine transitions at frequencies of \textcolor{ColorName1}{30.2} MHz and \textcolor{ColorName1}{43.3} MHz. In comparison, the $^{153}\mathrm{Eu}^{3+}$ ions show transitions at \textcolor{ColorName1}{77.1} MHz and \textcolor{ColorName1}{110.7} MHz for the same excited states. Furthermore, we determined the ground state hyperfine splittings by examining the frequency differences between the side-holes or anti-holes and the central hole. For $^{151}\mathrm{Eu}^{3+}$  ions, these frequencies are \textcolor{ColorName1}{19.7} MHz and \textcolor{ColorName1}{21.6} MHz, while for $^{153}\mathrm{Eu}^{3+}$ ions, they are \textcolor{ColorName1}{50.0} MHz and \textcolor{ColorName1}{55.2} MHz. 

\begin{figure}
	\centering
	\includegraphics[width=0.48\textwidth]{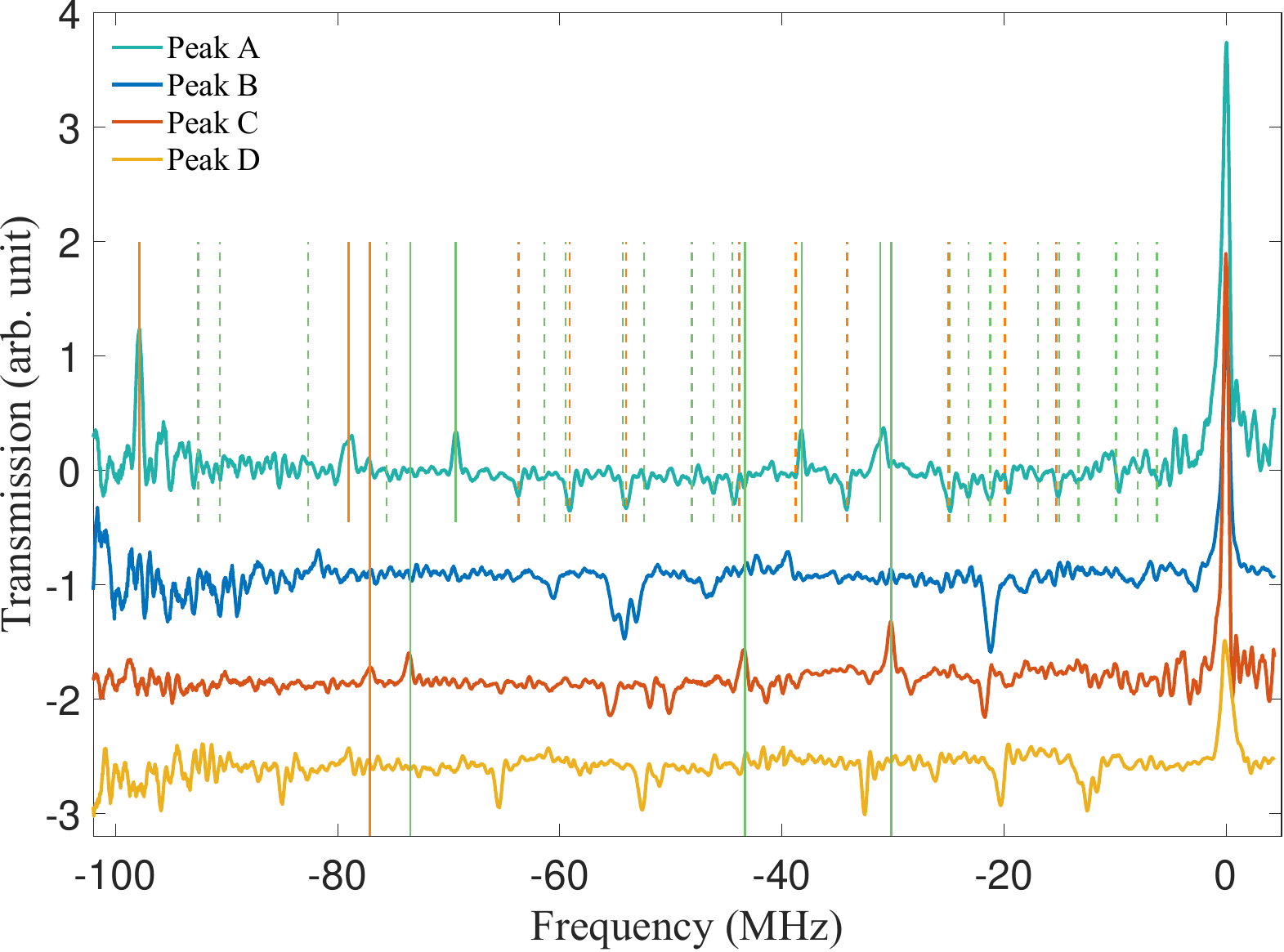}
	\caption{Under zero magnetic field conditions, the schematic diagram of the single-frequency \textcolor{ColorName1}{SHB} for \textcolor{ColorName1}{Peak} A, B, C and D is presented. \textcolor{ColorName1}{The figure clearly demonstrates that the} positions of \textcolor{ColorName1}{holes and anti-holes} for the four peaks are not consistent. Due to the limited number of effective holes and \textcolor{ColorName1}{anti-holes} in \textcolor{ColorName1}{Peak} B and D, \textcolor{ColorName1}{calculating their hyperfine transitions is not feasible. Only the calculated hyperfine transitions for Peak A are displayed.} The green and orange lines \textcolor{ColorName1}{correspond to the theoretical single-frequency \textcolor{ColorName1}{SHB} spectra of $^{151}\mathrm{Eu}^{3+}$ and $^{153}\mathrm{Eu}^{3+}$, respectively, with dashed lines indicating anti-hole positions and solid lines marking hole positions.}}
	\label{Fig:5}
\end{figure}

\textcolor{ColorName1}{However, the experiments conducted on other peaks reveal significant differences in both side-holes and anti-holes, as shown in Fig.\ref{Fig:5}. This suggests the hyperfine transitions among the four peaks differ for both $^{151}\mathrm{Eu}^{3+}$ and $^{153}\mathrm{Eu}^{3+}$  ions. For Peak A, the measured transition frequencies of excited hyperfine state are 31.2 MHz and 38.2 MHz for $^{151}\mathrm{Eu}^{3+}$  ions, while for $^{153}\mathrm{Eu}^{3+}$  ions, they are 79.0 MHz and 97.9 MHz. For the ground state, the measured hyperfine splitting for $^{151}\mathrm{Eu}^{3+}$  ions are 21.3 MHz and 23.2 MHz, and 54.0 MHz and 59.1 MHz for $^{153}\mathrm{Eu}^{3+}$  ions. A comparison of Peak A and C shows marked differences in both ground and excited state hyperfine transition frequencies between the two europium isotopes. }The maximum frequency difference between these peaks can reach up to \textcolor{ColorName1}{12.8} MHz. 

Based on these observations, we conclude that the hyperfine transitions among the four peaks \textcolor{ColorName1}{exhibit} inconsistent. \textcolor{ColorName1}{This phenomenon can be attributed to the charge imbalance between $\mathrm{Eu^{3+}}$ and $\mathrm{Ca^{2+}}$. Generally, \textcolor{ColorName1}{two charge compensation mechanisms are possible:} the first is to use alkali ions such as $\mathrm{Na}^{+}$ for compensation ($2\mathrm{Ca}^{2+} = \mathrm{Eu}^{3+}+\mathrm{Na}^{+}$), and the second employs a charge compensation mechanism consisting of 3$\mathrm{Ca}^{2+}$ and 2$\mathrm{Eu}^{3+}$ ($3\mathrm{Ca}^{2+} = 2\mathrm{Eu}^{3+}+\Box$, $\Box$ is a Ca site vacancy) \cite{10.2113/gsecongeo.94.3.423}. \textcolor{ColorName1}{Since $\mathrm{Na}^{+}$ was not co-doped during crystal growth, the second mechanism was considered dominant in this work.} Due to the charge compensation effect, the difference in oxidation states between $\mathrm{Eu}^{3+}$ and $\mathrm{Ca}^{2+}$ leads to lattice distortion. \textcolor{ColorName1}{Additionally, the O-W bond length may vary}, altering the local crystal field environment and causing the site symmetry of $\mathrm{Eu}^{3+}$ to deviate from $\mathrm{S}_4$ symmetry \cite{Su2008, C5DT00022J}.}

\begin{figure*}[h]
	\centering
	\includegraphics[width=1.0\textwidth]{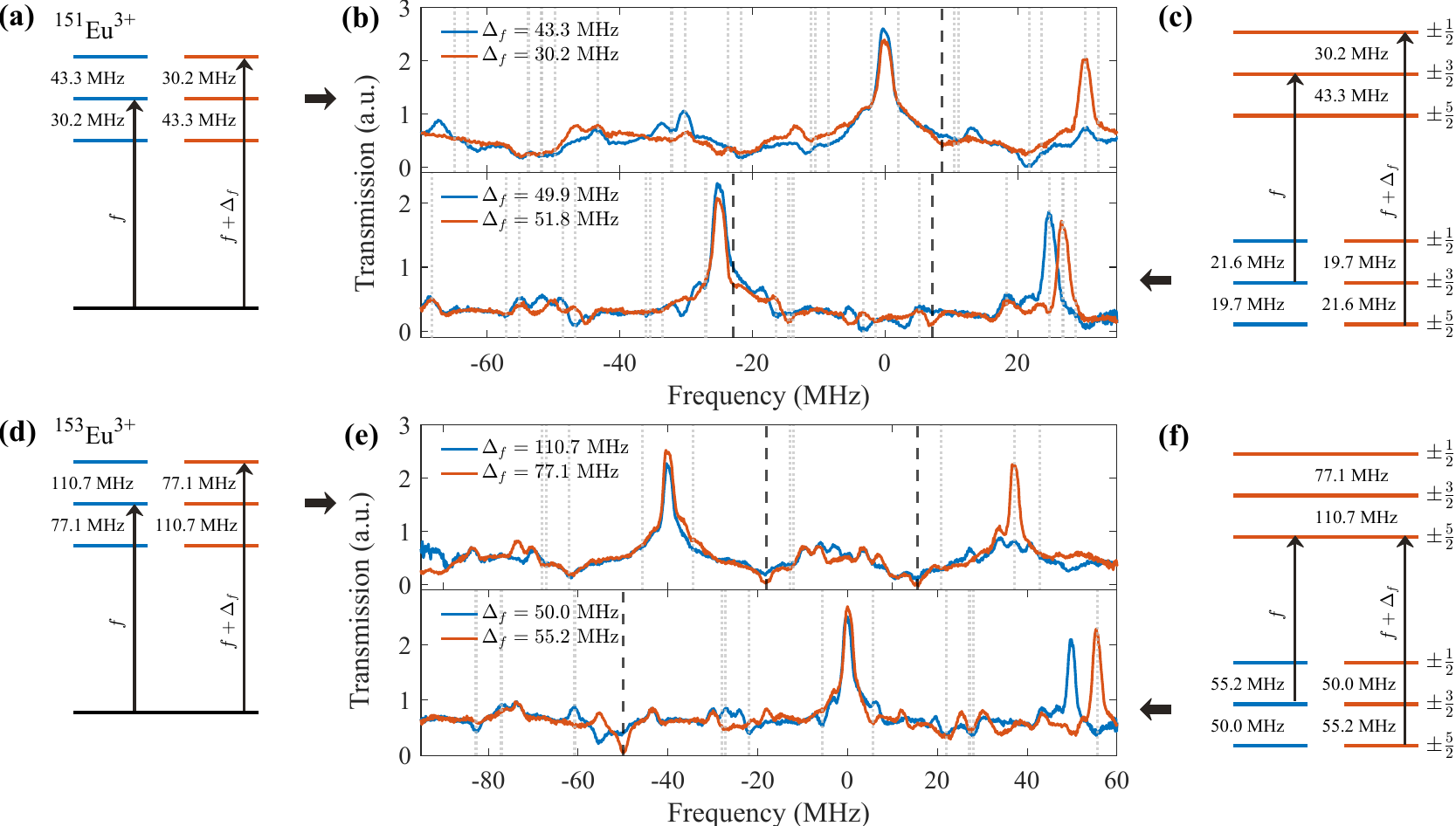}
	\caption{The dual-frequency \textcolor{ColorName1}{SHB} technique was used to determine the order of hyperfine energy levels for \textcolor{ColorName1}{both} ground and excited states. \textcolor{ColorName1}{Panels} (a), (c), (d), and (f) display \textcolor{ColorName1}{two possible hyperfine energy level configurations with their corresponding pump laser pairs; while panels (b) and (e) illustrate the hole spectra generated by these pump pairs.} The red spectrum shows an enhanced anti-hole depth at specific positions, \textcolor{ColorName1}{confirming that the red hyperfine structures in panels (a), (c), (d), and (f) represent the correct configurations. }The gray dotted lines in (b) and (e) \textcolor{ColorName1}{indicate anti-holes from single-frequency SHB, which serve as references to identify enhanced anti-holes (marked by black dashed lines).} Specifically, panels (a) and (c) represent the possible hyperfine energy level structures for the excited and ground states of $^{151}\mathrm{Eu}^{3+}$, \textcolor{ColorName1}{with the red configuration in (c) identified as the correct structure.} Likewise, panels (d) and (f) depict the possible hyperfine energy level structures for the excited and ground states of $^{153}\mathrm{Eu}^{3+}$, \textcolor{ColorName1}{where the red configuration in (f) is validated as the correct structure.}}
	\label{Fig:6}
\end{figure*}

\textcolor{ColorName1}{We then determined} the order of hyperfine energy levels of the relevant ground and excited states of $\mathrm{Eu^{3+}}$ in Peak C by using dual-frequency \textcolor{ColorName1}{SHB}  technique \cite{nilsson2004hole,lauritzen2012spectroscopic,jin2022faithful}. \textcolor{ColorName1}{In this method, two laser beams with distinct frequencies irradiate the sample simultaneously, thereby breaking} the spectral symmetry. When the frequency difference between the lasers matches the hyperfine splitting \textcolor{ColorName1}{energy}, the intensity of \textcolor{ColorName1}{anti-holes} at specific positions \textcolor{ColorName1}{is} enhanced. The results are shown in Fig.\ref{Fig:6}\textcolor{ColorName1}{. For comparison, all anti-holes from} single-frequency \textcolor{ColorName1}{SHB} are plotted \textcolor{ColorName1}{with gray dotted lines, while those enhanced during dual-frequency SHB are marked with black dashed lines to identify the anti-holes used for determining the hyperfine energy level order.}  \textcolor{ColorName1}{Although weak absorption limited the detection to only one or two enhanced anti-holes, the hyperfine structures of $^{151}\mathrm{Eu^{3+}}$ and $^{153}\mathrm{Eu^{3+}}$ can be determined, as depicted in Fig.\ref{Fig:6} red part}. 

 \textcolor{ColorName1}{To determine the }excited state energy level order of $^{151}\mathrm{Eu}^{3+}$ ions, we \textcolor{ColorName1}{employed} two sets of pump laser frequency combinations: $f$ and $f + \Delta_f$, \textcolor{ColorName1}{where} $\Delta_f = $ \textcolor{ColorName1}{30.2} MHz (\textcolor{ColorName1}{43.3} MHz) as shown in Fig.\ref{Fig:6}a red part (blue part). The experimental results indicate a significant enhancement in the \textcolor{ColorName1}{anti-hole intensity} at \textcolor{ColorName1}{8.6} MHz in the spectrum (black dashed line in the top of Fig.\ref{Fig:6}b), suggesting that the hyperfine structure of the excited state \textcolor{ColorName1}{corresponds to the configuration illustrated} in Fig.\ref{Fig:6}a red part. 
Subsequently, \textcolor{ColorName1}{to determine the ground state hyperfine energy level order} of $^{151}\mathrm{Eu}^{3+}$ ions, we \textcolor{ColorName1}{tested two additional pump laser frequency combinations:} $f$ and $f + \Delta_f$, $\Delta_f = $ \textcolor{ColorName1}{51.8} MHz (\textcolor{ColorName1}{49.8} MHz) as shown in Fig.\ref{Fig:6}c red part (blue part). Utilizing these \textcolor{ColorName1}{combinations, we observed a pronounced increase in anti-hole intensity} (black dashed line in the bottom of Fig.\ref{Fig:6}b)\textcolor{ColorName1}{, demonstrating that the} $f$ and \textcolor{ColorName1}{$f + 51.8$} MHz \textcolor{ColorName1}{combination} can significantly enhance the intensity of the anti-holes. \textcolor{ColorName1}{This confirms that the }hyperfine structure of the ground state \textcolor{ColorName1}{matches the model} in Fig.\ref{Fig:6}c red part. \textcolor{ColorName1}{For} the $^{153}\mathrm{Eu}^{3+}$ ions, the same methodology was employed. The results shown in Fig.\ref{Fig:6}f red part indicate the correct order of the hyperfine structure \textcolor{ColorName1}{for both} ground \textcolor{ColorName1}{state} and excited state. 

The effective quad\-ru\-pole Hamiltonian of the $\mathrm{Eu^{3+}}$ system primarily determines the hyperfine structure of both the ground state and the excited state under zero magnetic field conditions. It can be expressed as \cite{MACFARLANE198751}
\begin{flalign}
	&H_{Q}=D(I_{z}^{2}-I(I+1)/3)+E(I_x^2-I_y^2) &
\end{flalign}
where D and E corresponds to the combined quadrupole \textcolor{ColorName1}{constant} and second-order hyperfine coupling \textcolor{ColorName1}{constant} along their respective principal axes, respectively. Table.\ref{Tabel:2} \textcolor{ColorName1}{lists the} parameters D and E \textcolor{ColorName1}{obtained} from the fitting for the ground and excited states of two isotopes in Peak C. Here, the sign of E does not \textcolor{ColorName1}{affect} the order of the hyperfine energy levels\textcolor{ColorName1}{; therefore, its absolute value is reported.} Meanwhile, the splitting ratio of two isotopes for Peak C, corresponding to \textcolor{ColorName1}{the nuclear quadrupole moment ratio} $ Q({}^{153}\mathrm{Eu})/Q({}^{151}\mathrm{Eu})$, \textcolor{ColorName1}{is found to be approximately 2.55}. This is consistent well with previous report \cite{babbitt1989optical}.

\begin{table}[h]
	\centering
	\caption{In $^{151}\mathrm{Eu}$ and $^{153}\mathrm{Eu}$, the hyperfine transitions in optical ground states($\nu_{g1}$ and $\nu_{g2}$) and optical excited states($\nu_{e1}$ and $\nu_{e2}$), along with their corresponding effective quadrupole spin Hamiltonian parameters (D and E).}
	\label{table4}
	\begin{tabular}{l ccc}
		\toprule
		 parameter & $^{151}\mathrm{Eu}$ & $^{153}\mathrm{Eu}$ & $^{153}\mathrm{Eu}/^{151}\mathrm{Eu}$  \\
		\midrule
		 $\nu_{g1}$ (MHz) & \textcolor{ColorName1}{19.7} & \textcolor{ColorName1}{50.0} & 2.54 \\
		 $\nu_{g2}$ (MHz) & \textcolor{ColorName1}{21.6} & \textcolor{ColorName1}{55.2} & 2.55 \\
		 $\nu_{e1}$ (MHz) & \textcolor{ColorName1}{30.2} & \textcolor{ColorName1}{77.1} & 2.56 \\
		 $\nu_{e2}$ (MHz) & \textcolor{ColorName1}{43.3} & \textcolor{ColorName1}{110.7} & 2.56 \\
		 $D_g$ (MHz) & \textcolor{ColorName1}{-6.0} & \textcolor{ColorName1}{-15.3} & 2.55 \\
		 $|E_g|$ (MHz) & \textcolor{ColorName1}{1.8} & \textcolor{ColorName1}{4.5} & 2.53 \\
		 $D_e$ (MHz) & \textcolor{ColorName1}{-11.5} & \textcolor{ColorName1}{-29.3} & 2.56 \\
		 $|E_e|$ (MHz) & \textcolor{ColorName1}{2.2} & \textcolor{ColorName1}{5.7} & 2.56 \\	
		\bottomrule
	\end{tabular}
	\label{Tabel:2}
\end{table}

\subsection{Lifetimes of $\mathrm{{}^5 D_0}$ excited state and $\mathrm{{}^7 F_0}$ ground hypferfine state}

The state \textcolor{ColorName1}{lifetime} of $\mathrm{Eu}^{3+}$ is also \textcolor{ColorName1}{a} critical \textcolor{ColorName1}{parameter} for evaluating the properties of materials, since \textcolor{ColorName1}{it provides the upper limit} of the optical and spin coherence times T$_2$ of the system (T$_2 \leqslant$ 2T$_1$). In this work, we measured the $\mathrm{{}^5 D_0}$ state lifetime of four peaks by collecting the fluorescence \textcolor{ColorName1}{decay} of the ${}^5\mathrm{D}_0\leftrightarrow{}^7\mathrm{F}_2$ transition and the $\mathrm{{}^7 F_0}$ ground \textcolor{ColorName1}{hyperfine} state lifetimes \textcolor{ColorName1}{using the SHB} technique. In the later experiments, we first burnt a hole for \SI{100}{ms} with a laser power of 15 mW at the center of the absorption line. Then we probed the hole depth, which indicates \textcolor{ColorName1}{that} the population redistribution occurs among three hyperfine ground state levels, with a laser pulse of 1 mW power. The results are listed in Table \ref{Table:1}.
	
\textcolor{ColorName1}{The fluorescence lifetimes (T$_1$) of the four peaks differ slightly but are similar to those observed previously in $\mathrm{Eu^{3+}}$:$\mathrm{CaWO_4}$} crystals \cite{zhang2019structure,xiao2024eu3+}. \textcolor{ColorName1}{Additionally, Peak B exhibits the longest value at $ 597 \pm 2~\mu$s, which aligns with its longest optical coherence time T$_2=114 \pm 7~\mu$s}. However, \textcolor{ColorName1}{the spectral hole lifetimes (T$_{1,spin}$)} of \textcolor{ColorName1}{the four peaks exhibit significant variations.} \textcolor{ColorName1}{Specifically, T$_{1,spin}$ values for Peak B ($1161 \pm 3$ s) and Peak D ($1215 \pm 4$ s) are substantially longer than those for Peak A ($688 \pm 1$ s) and Peak C ($779 \pm 9$ s).} \textcolor{ColorName1}{Although Peak B has} a comparable absorption depth \textcolor{ColorName1}{to Peak} C, its longer excited state lifetime T$_1$ and ground hyperfine state lifetime T$_{1,spin}$ \textcolor{ColorName1}{enable} longer storage times than other peaks.

\section{Conclusions}
In our study, we observed four distinct ${}^7\mathrm{F}_0\leftrightarrow{}^5\mathrm{D}_0$ optical transitions in \textcolor{ColorName1}{$\mathrm{Eu^{3+}}$:$\mathrm{CaWO_4}$}  crystals by analyzing their fluorescence and absorption spectra. We believe these transitions are primarily due to \textcolor{ColorName1}{the} charge compensation effect. To further investigate these transitions, we used \textcolor{ColorName1}{the SHB}  technique to analyze the hyperfine transitions of \textcolor{ColorName1}{the} four peaks and identified inconsistencies among them. Specifically, we detailed the hyperfine energy level structure of \textcolor{ColorName1}{Peak} C and measured the spectral hole lifetimes for all four peaks. \textcolor{ColorName1}{Our results showed that the spectral hole lifetimes extend up to 10 minutes, and the optical coherence time can reach 100 microseconds}. These findings provide essential experimental evidence for understanding the spectral characteristics of \textcolor{ColorName1}{$\mathrm{Eu^{3+}}$:$\mathrm{CaWO_4}$}  crystals, advancing our knowledge in this area.

\section{Acknowledgments}
We thank Prof. H. Chen and Prof. Z. Sun for kindly growing the crystal samples used in this work. This work was supported by the Innovation Program for Quantum Science and Technology (No.2021ZD0301204), the National Natural Science Foundation of China (Grant Nos.12304454, 12004168, 11904159), National Key Research and Development Program of China
(Grant No.2022YFB3605800), Guangdong Innovative and Entrepreneurial Research Team Program (Grant No. 2019ZT08X324), Guangdong Basic and Applied Basic Research Foundation (Grant No.2021A1515110191), the Key-Area Research and Development Program of Guangdong Province (Grant No.2018B030326001) and The Science, Technology and Innovation Commission of Shenzhen Municipality (KQTD202108110900-49034).

\bibliographystyle{unsrt}
\bibliography{bib/cas-refs}

\end{document}